\documentstyle[11pt]{article}
\newcommand{\al}{\alpha}
\newcommand{\si}{\sigma}
\newcommand{\om}{\omega}
\newcommand{\fee}{\varphi}

\newcommand{\pa}{\partial}
\newcommand{\be}{\begin{equation}}

\newcommand{\ee}{\end{equation}}
\newcommand{\bea}{\begin{eqnarray}}
\newcommand{\eea}{\end{eqnarray}}
\begin{document}

\begin{flushright}
PSI-PR-95-09\\
IPS Report 95-14\\
UCY-PHY-95/5\\[2cm]
\end{flushright}
\begin{center}
{\bf{\Large Full QCD with the L\"uscher local bosonic action}}\\[1cm]
 C.Alexandrou $^{1,2}$, A. Borrelli $^{2}$, Ph. de Forcrand $^3$,
 A. Galli $^2$, F. Jegerlehner $^2$\\[0.5cm]
{\it  $^1$ Department of Natural Sciences, University of Cyprus,
CY-1678 Nicosia, Cyprus\\
$^2$ Paul Scherrer Institute, CH-5232 Villigen, Switzerland\\
$^3$ IPS, ETH-Zentrum, CH-8092 Z\"urich, Switzerland\\}
\end{center}

\medskip
\begin{center}
{\bf Abstract}\\[1cm]
\end{center}
We investigate L\"uscher's method of including dynamical Wilson fermions
in a lattice simulation of QCD with two quark flavours.
  We measure the accuracy of the
 approximation by comparing it with Hybrid Monte Carlo results for gauge
 plaquette and Wilson loops.
We also introduce an additional global Metropolis step in the update.
We show that the complexity of L\"uscher's algorithm
 compares favourably with that of the Hybrid Monte Carlo.
\newpage

\section{Introduction}

Taking into account the effects of sea-quarks in lattice QCD simulations
presents serious te\-chni\-cal difficulties. The main problem is the
excessive cost in computer time due to the need of evaluating the fermion
matrix determinant for each gauge field update step.  The most used
algorithm, Hybrid Monte Carlo (HMC) \cite{hmc},
  encounters dramatic problems of
CPU time costs and slowing down in the limit of small quark masses,  a fact
that has generally forced people to use the valence (quenched)
approximation in which the fermion determinant is assumed to be unity.  For
many observables the results obtained in this approximation fit quite well
with phenomenologically known properties of QCD,
but for other observables
the quenched approximation is definitely too crude (for example $\alpha_s$).\\

The urgent need to be able to perform calculations in full QCD has
motivated substantial activity to overcome the technical problems caused by
the full dynamical treatment of the fermions. The search for a less
expensive algorithm for lattice sea-quarks simulation is therefore a very
actual subject, on which recently a new proposal has been made by
M. L\"uscher \cite{luescher}. The new method consists in approximating the
inverse of the fermion  matrix with a polynomial, and then interpreting the
determinant  as the partition function of a local bosonic system.
Simulating this bosonic system will enable us to evaluate the effects of
the  sea-quarks.\\

This algorithm is quite general, and its application to various
systems has been and is being studied, with the aim of optimizing it and
rendering it economically viable. In this paper, we will report the
results of a first exploratory  application of L\"uscher's algorithm
to an SU(3) Lagrangian with two degenerate flavours of Wilson quarks.
The observables measured are the gauge plaquette and Wilson loops,
for which we study the autocorrelation time and precision of the approximation
for numerous sets of parameters.
We also show the effect of inserting
a global Metropolis step in the update,
as suggested in reference \cite{peardon},
with the aim of making the algorithm more precise and more efficient.

\section{The algorithm}

The partition function of lattice QCD with two fermion flavours
can be written as
\be
Z=\int [dU] {\rm det} Q^2 e^{-S_{G}[U]}
\ee
where $S_{G}[U]$ is the pure gauge action and the fermion matrix $Q$ has the
form:
\be
Q=\frac{\gamma_5(D+m)}{c_M~M},~~~~~~ {\rm with}~~~~~~M=\frac{8}{a}+m.
\ee
Here $a$ is the lattice spacing, and
 $Q$  has been chosen hermitian, and with eigenvalues
 contained in the interval $(-1,1)$. The constant $c_M\geq 1$ is fixed
to the values $1.1$ as explained in ref. \cite{luescher}.\\

As anticipated, the first step is  approximating $1/Q^2$ with
 a polynomial $P(Q^2)$. In choosing the form of $P(s)$,
 we follow \cite{luescher},
 where its explicit expression
in terms of Chebyshev polynomials is given.

$P(s)$ is of even degree $n$, and
approximates the
function $1/s$ in an interval $\epsilon \le s \le 1$, where $\epsilon$
is a parameter to be chosen according to the range
of eigenvalues of $Q^2$; the
  $n$ roots of $P(s)$, $z_k$,
lie on an ellipse in the complex plane with foci $\epsilon$ and $1$,
and minor axis of length $2\sqrt{\epsilon}$;
 they come in complex conjugate pairs, with
 $Im\,\,z_k\neq 0$.
Since the roots come in complex conjugate pairs, $P(s)$ can be written
in the form:

\be
P(s)=\mbox{const}\times\prod_{i=1}^{n}(s-z_k)=
\mbox{const}\times\prod_{i=1}^{n}(\sqrt{s}-\sqrt{\bar z_k})
(\sqrt{s}-\sqrt{z_k})~.
\ee
A measure of the relative error
of the approximation is given by
 $R(s)=P(s)s-1$ , which in the interval $\epsilon\leq s\leq 1$
 is bounded by
\be
|R(s)|\leq
2\left(\frac{1-\sqrt{\epsilon}}{1+\sqrt{\epsilon}}\right)^{n+1} \equiv
 ~~ \delta (n,\epsilon)\label{delta}~.
\ee
When $s<\epsilon$ the polynomial continues to converge, but with an
$s$-dependent exponential rate approaching zero in the $s\rightarrow
0$ limit.\\

We can now introduce a set of boson fields $\phi_k$ $(k=1,...n)$
with partition function $Z_b$, approx\-imating the fermion determinant
\be
Z_b[U]=\int
\prod_{k=1}^n[d\phi_k][d\phi_k^\dagger]e^{-
\phi^\dagger_k(Q-\sqrt{\bar z_k})(Q-\sqrt{z_k})\phi_k }
=
\frac{1}{detP(Q^2)}
\simeq
detQ^2 ~.
\label{lagr}
\ee
The partition function of full QCD is then approximated by the local
bosonic action
\be
Z=\int[dU] \> detQ^2e^{-S_{G}[U]}\> \simeq\int[dU]\> Z_b[U]\> e^{-S_{G}[U]}~.
\label{approx}
\ee
Making use of this action,
we may now simulate the fermion determinant by locally
updating the boson fields $U$ and $\phi$,
 using algorithms like heat-bath and
over-relaxation. The implementation of the program is described in
Appendix A.

\section{Numerical simulations on the $4^4$ lattice for the pure L\"uscher
algorithm}

The numerical simulations of the theory described by (\ref{approx})
 were
performed on a $4^4$ lattice
 for various choices of the parameters $\epsilon$ and $n$, at $\beta=6.0$,
  for $\kappa=0.12 ~~{\rm and}~~ 0.14$ (in the deconfined phase),
and at $\beta=5.0$ for $\kappa=0.15$ (in the confined phase).\\

For each set of parameters, an average of 10,000 sweeps were performed,
each sweep consisting of one $\phi$ heat-bath, one $\phi$ over-relaxation,
 and one to six $U$ over-relaxations (see further on for discussion),
 ergodicity being ensured by the $\phi$ heat-bath. After each
 sweep the gauge boson plaquette was measured.

For all $\kappa$ values, the minimum
 eigenvalue of $Q^2$ was measured for approximately 10,000 configurations.
The average results are:
\bea
       && 0.015 \pm 0.001 ~~~~~~~~~~      (\kappa=0.12,~ \beta=6) \\\nonumber
       && 0.006 \pm 0.001 ~~~~~~~~~~      (\kappa=0.14,~ \beta=6)\\\nonumber
       && 0.003 \pm 0.001 ~~~~~~~~~~      (\kappa=0.15,~ \beta=5)\nonumber
\eea
The averages of the lowest eigenvalues are shown in Fig. 1
for $\kappa=0.12$ and $\beta=6$.
As a check, measurements of the plaquette with
different numbers of $U$ field over-relaxation steps have been done,
monitoring the autocorrelation time: no variation was observed,
indicating that the autocorrelation time was dominated by the dynamics of
the $\phi$ fields.
 Consequently we chose to perform in subsequent runs
 only one $U$ over-relaxation per sweep.\\

For comparison, the plaquette was measured for the same lattice parameters
and $\kappa$ values also using a HMC
 program. The results of all the simulations are reported in Table 1.
In Fig. 2 the plaquette values are plotted as a function of
 $\delta(n,\epsilon)$.\\

As can be seen from Table 1, many values of $\epsilon$ chosen
are quite high with respect to the lowest $Q^2$ eigenvalue;
despite that, the agreement with the HMC results remains good,
on this $4^4$ lattice, up to $\epsilon$  values ten times
larger than the lowest $Q^2$ eigenvalue.
This shows that the first few low-energy modes of the fermion determinant
do not couple strongly to the plaquette. On larger lattices where
non-local observables can be measured better, one would expect
 that $\epsilon$
will have to be chosen closer to the lowest eigenvalue.\\

It is also interesting to observe that the convergence of the
plaquette to its correct value is not exponential in $n$; in fact it
is not necessarily monotonic. For $\beta=5.$ for example it approaches
the correct value from above, while the quenched ($n=0$) value is
below the full QCD one. This unexpected behaviour occurs when
$\epsilon$ is taken much larger then $\lambda_{min}(Q^2)$. In such
cases $\delta$, taken alone, can not be a good measure of the error:
The bulk of the error comes from small eigenvalues
$\lambda<\epsilon$.\\

For $\kappa=0.12$, the results show that we can push $\epsilon$ up to
0.1, while we see a definite breakdown of the approximation at
 $\epsilon = 0.5,~~n=4 $ ($\delta \sim 2. \times 10^{-4}$).
 As we approach more critical $k$ values,
 the eigenvalues of $Q^2$ decrease and, for $\kappa=0.14$,
 we see that the approximation breaks down already at $\epsilon=0.08,
{}~~n=16$, ($\delta \sim 1. \times 10^{-4}$).\\

For $\beta=5$ and $\kappa=0.15$, in the confined phase,
the
minimum eigenvalue is $\sim 0.003$, and
satisfactory results can be obtained
at $\epsilon=0.01$, $n=50$,  ($\delta \sim 7 \times 10^{-5}$). \\

\begin{table}
\begin{center}
\begin{tabular}{|c|l|l|l|l|l|l|l|}\hline
$\beta$ & $\kappa$ & $\epsilon$ & $n$ & plaquette & $\tau$ &
$\delta(n,\epsilon)$ & $n/\sqrt{\epsilon}$\\\hline\hline
 &&&&&&&\\
6.0 & 0.12 & 0.005 & 20 & 0.6108(5) & 8  &   0.1     & $\sim$ 283 \\
 &    &       & 60 & 0.6016(6) & 25 & 3.5$\times 10^{-4}$ & $\sim$ 849 \\
 &     & 0.01  & 40 & 0.6018(4) & 10 & 5.3$\times 10^{-4}$ & ~~ 400 \\
 &     &       & 70 & 0.6023(7) & 23 & 1.3$\times 10^{-6}$ & ~~ 700 \\
 &     & 0.016 & 40 & 0.6020(4) & 8  & 6.0$\times 10^{-5}$ & $\sim$ 316 \\
 &     &       & 70 & 0.6017(4) & 13 & 2.0$\times 10^{-8}$ & $\sim$ 553 \\
 &     & 0.04  & 30 & 0.6020(4) & 5  & 7.0$\times 10^{-6}$ & ~~ 150 \\
 &     &       & 50 & 0.6021(4) & 8  & 2.1$\times 10^{-9}$ & ~~ 250 \\
 &     &       & 70 & 0.6023(5) & 9  & 6.0$\times 10^{-13}$ & ~~ 350 \\
 &     & 0.08  & 14 & 0.6024(2) & 3  & 3.2$\times 10^{-4}$ & $\sim$ 49 \\
 &     &       & 50 & 0.6016(3) & 6  & 2.6$\times 10^{-13}$ & $\sim$ 177 \\
 &     & 0.5   &  4 & 0.5966(3) & 2  & 2.9$\times 10^{-4}$ & $\sim$ 6 \\
 &     &       &  6 & 0.5968(5) & 11 & 8.7$\times 10^{-6}$ & $\sim$ 8 \\
 &     &       &  8 & 0.5984(4) & 12 & 2.6$\times 10^{-7}$ & $\sim$ 11 \\
 &HMC  &       &    & 0.6026(3) &  4  &                      &    \\
 &&&&&&&\\
 &0.14 & 0.02  & 50 & 0.6089(4) & 9  & 1.0$\times 10^{-6}$ & $\sim$ 354 \\
 &     &       & 90 & 0.6075(6) & 20 & 1.0$\times 10^{-11}$ & $\sim$ 636 \\
 &     & 0.04  & 30 & 0.6084(3) & 5  & 7.0$\times 10^{-6}$ & ~~ 150 \\
 &     &       & 60 & 0.6089(5) & 8  & 3.6$\times 10^{-11}$ & ~~ 300 \\
 &     &       & 90 & 0.6082(7) & 14 & 1.9$\times 10^{-16}$ & ~~ 450 \\
 &     & 0.08  & 16 & 0.6062(3) & 4  & 1.0$\times 10^{-4}$ & $\sim$ 56 \\
 & HMC  &       &    & 0.6094(2) & 5   &                      &    \\
 &&&&&&&\\
5.0 & 0.15 & 0.005  & 50 & 0.413(1) & *  & 1.0$\times 10^{-3}$  & $\sim$ 707 \\
 &     &       & 80 & 0.413(2) & * & 2.0$\times 10^{-5}$  & $\sim$ 1131 \\
 &     & 0.01  & 50 & 0.415(1) & * & 7.0$\times 10^{-5}$  & ~~ 500 \\
 &     &       & 80 & 0.415(1) & * & 2.0$\times 10^{-7}$  & ~~ 800 \\
 &     & 0.05  &  6 & 0.4239(5) & 14 & $ 8.2\times 10^{-2}$  & $\sim$ 27 \\
  &     &     & 14 & 0.4180(7) & 50 & 2.2$\times 10^{-3}$  & $\sim$ 63 \\
  &     &     & 50 & 0.4162(12) & * & 2.0$\times 10^{-10}$  & $\sim$ 224 \\
&     &       & 80 & 0.4163(10) & * & 2.0$\times 10^{-16}$  & $\sim$ 358 \\
 & HMC  &       &    & 0.4155(8) & 7   &               &    \\\hline
\end{tabular}
\caption{Simulation results for the plaquette on a $4^4$ lattice at
$\beta=6$ for the L\"uscher algorithm and the HMC algorithm. Each
plaquette value is an average of about $10'000$ measurements.
A star $*$ means that the autocorrelation time was too long to be
measured accurately.
The autocorrelation time of the L\"uscher algorithm is measured in
unit of iteration sweeps, consisting of one HB and OR of the $\phi$
and one OR of the $U$.
The autocorrelation time of the HMC is measured in unit of
trajectories, each of them consisting of 12 molecular dynamic steps.}
\end{center}
\end{table}

Since the CPU costs are expected to be at least
proportional to the number of fields $n$,
and, in turn, a smaller value of $\epsilon$ requires
a larger $n$,
 we conclude that the most economical
 program will use the highest
$\epsilon$
and lowest $n$ values compatible with a good approximation of
the fermion determinant.
A more detailed discussion of the CPU costs for obtaining
a set of uncorrelated configurations is found in the
next section.
 From our observation of the plaquette on a $4^4$ lattice,
it appears that one can raise $\epsilon$
up to values significantly larger than the lowest
$Q^2$ eigenvalue.\\

It will be shown in section 5,
that the situation can be further improved by the insertion
of a global Metropolis test in the updating procedure for the $\phi$
and $U$ fields. The modified method
 will allow us to maintain a good approximation for
higher $\epsilon$ and therefore lower $n$, thus
reducing the CPU cost of the simulation.

\section{Autocorrelation and scaling behaviour for the pure L\"uscher
algorithm}

 We have measured the autocorrelation of the plaquette in the
deconfined and confined phases.
In Fig. 3, the plaquette autocorrelation time $\tau$
is shown versus the ratio $\frac{n}{\sqrt{\epsilon}}$, in the
deconfined phase: the relationship
between the two appears to be linear.
On the other hand, if we freeze the bosonic fields $\phi$, the
autocorrelation time of the plaquette is ${\cal O}(1)$ sweep
of the gauge fields.
So the long correlation times are caused by the slow dynamics of
the $\phi$ fields. These bosons
are updated by local algorithms (e.g. heat-bath) which have a critical
dynamical exponent of 2 in the worst case scenario. This means that, if the
correlation length of
the bosonic action is $\xi$ in lattice units, then the correlation time
will be $\sim \xi^2$. What then is the correlation length ?
The bosonic action is of the form
$|(Q-\mu_k)\phi_k|^2+|\nu_k\phi_k|^2 =
|(\gamma_5 Q -\gamma_5\mu_k)\phi_k|^2+ |\nu_k\phi_k|^2$.
In the deconfined phase the Dirac matrix $\gamma_5 Q$ is chirally symmetric.
But the second term $\gamma_5 \mu_k$ is not: it breaks chiral symmetry,
with a characteristic length scale $1/\mu_k$.
Therefore we expect that the smallest mass term $\mu=min_k(\mu_k)$ will
control the autocorrelation time: $\tau\sim\xi^2\sim\frac{1}{\mu^2}$.
It is easy to observe from the definition of the roots that the
smallest $\mu^{-2}$ is proportional to $n/\sqrt{\epsilon}$.
The same conclusion $\tau \sim n/\sqrt{\epsilon}$ can be reached also
by considering the correlation length $1/\nu_k$ given by the third term
of the bosonic action above \cite{luescher}.\\

In the confined phase, an additional length scale $1/m_\pi$ appears
because of chiral symmetry breaking. This length scale may obscure the
previous relationship $\tau \sim 1/\sqrt{\epsilon}$, and weaken the
dependence of $\tau$ on $\epsilon$. Unfortunately our numerical simulations
are not accurate enough to resolve this issue: they only indicate that
$\tau$ increases very much in the confined regime, for a given quark mass
and number of bosonic fields.\\

While the cost in memory of the program is proportional to the volume of the
lattice and to the number of auxiliary boson fields, the work (or CPU time)
necessary to obtain
two uncorrelated configurations is proportional to the
number of boson fields times the autocorrelation time $\tau$.

Assuming the same $1/\sqrt{\epsilon}$ dependence of $\tau$ in both phases,
which should be a worst case scenario,
we can predict the slowing down of the algorithm as the quark mass $m_q$
decreases. $\epsilon$ must be tuned according to the lowest eigenvalue
of the $Q^2$ matrix and is therefore proportional to $m_q^2$.
Requiring the error $\delta$ to be constant one obtains from (\ref{delta})
that $n \sim 1/m_q$ and therefore, for a fixed volume, the autocorrelation
time increases like $m_q^{-2}$. Since the work per sweep is proportional
to the number $n \sim 1/m_q$ of fields to update, one finally obtains that
the work to obtain a new, decorrelated configuration, scales like $m_q^{-3}$.

 As the volume $V$ of the lattice
is increased, the scaling behaviour of the algorithm
can be easily predicted from the exponential convergence
of the approximation eq. (\ref{delta}). Let us keep the quark mass, and
thus $\epsilon \sim m_q^{-2}$, fixed. Then requiring that the accumulated
error of the approximation remain constant as $V$ is increased leads to
$n \sim \log V$. The autocorrelation time is proportional to $n$, and the
work to obtain a new, decorrelated configuration proportional to $n^2$,
so that the work scales like $V (\log V)^2$.

To summarize this section, the needed work per new configuration is
at most
proportional to $V~(\log V)^2~ m_q^{-3}$. This compares favourably with that of
the
HMC algorithm, $V^{5/4}~m_q^{-7/2}$ or $V^{5/4}~m_q^{-13/4}$, depending
on the scaling of the trajectory length with $m_q$ \cite{gupta}.
Either way, the L\"uscher algorithm is asymptotically faster,
with respect to both large volumes and small quark masses.

\section{The systematic error and the exact algorithm}
The exact partition function of QCD is given by
\bea
Z&=&\int[dU] \> \frac{detQ^2 \> detP(Q^2)}{detP(Q^2)}
e^{-S_{G}[U]} \\ \nonumber
&=&\int[dU]\> det\left(Q^2P(Q^2)\right)Z_b[U]e^{-S_{G}[U]} \quad.
\label{exact}
\eea

The partition function used in eq. (\ref{approx}) corresponds to approximating
$det\left(Q^2P(Q^2)\right)$ with unity.
Hence, for a given configuration of gauge fields $U$, the error involved in the
approximation of the determinant is given by
\be
R[U]=detQ^2P(Q^2)-1=\prod_i\lambda_iP(\lambda_i)-1\label{ru}
\ee
where the last step follows by simultaneously diagonalising
$P(Q^2)$ and $Q$. $\lambda_i$ are the
 eigenvalues of $Q^2$ for the configuration
$U$.

The actual value of this systematic error
 changes between two different
configurations $U$ and $U'$ as $R[U]$ does.
The amplitude of these fluctuations can be tuned by an appropriate
choice of the approximation parameters, the polynomial degree $n$
and the cut-off $\epsilon$.
Of course, the cost in memory and CPU
time will dramatically increase if we are
 trying to improve the approximation.\\

A possibility to reduce the error, is to introduce the effect of
the term $R[U]+1$ in the simulation by the insertion of a Metropolis step
in the update process, according to the following
 scheme \cite{peardon}:\\[0.1cm]

\begin{itemize}
\item We start with a configuration ($U$, $\phi$);
the operator $R[U]+1$ is calculated for the configuration $U$.
\item The $\phi$ and $U$ fields are updated {\em m} times, by performing
a $\phi$ heat-bath together with over-relaxation for the ${\phi}$ fields
and over-relaxation for the  $U$ fields.
\item The error $R[U']+1$ is calculated for the new gauge configuration
$U'$.
\item The new configuration ($U'$, $\phi '$) is accepted
 with probability
$$P_{acc}=min\left[1,\frac{R[U']+1}{R[U]+1}\right]~.$$
\end{itemize}

This combination of heat-bath, over-relaxation and Metropolis update
 ensures that the Lagrangian simulated is the exact QCD one of eq.
 (\ref{exact}).\\

The computation of $R[U]+1$ is performed by computing the $Q^2$
eigenvalues $\lambda_i$ and using:
\be
R[U]+1=\prod_{i=1}^L \lambda_iP(\lambda_i) \quad.
\label{ll}
\ee
 This is a
computationally intensive step, because in theory the full
 spectrum of the matrix $Q^2$
has to be determined. In practice, however, when the
polynomial is precise enough ($\delta(n,\epsilon)<10^{-4}$),
 the dominant contribution
to (\ref{ll}) comes from the low-lying eigenvalues $\lambda<\epsilon$
of $Q^2$, and the product (\ref{ll}) can be cut-off at some lower
bound $L'$. With this method the CPU cost invested for the evaluation of
$R[U]$ becomes irrelevant if the Metropolis tests are divided by a sufficient
number {\em m} of
updating steps
 and the number of eigenvalues $\lambda<\epsilon$ is not
excessive. On a $4^4$ lattice, possibly thanks to
 the low density of the eigenvalues,
 we can choose $\epsilon$ even ten times larger than the lowest
eigenvalue, reducing even further the CPU time.
Of course, on bigger lattices, $\epsilon$ will have to be chosen closer to
the lowest eigenvalue. Notice that the work needed for the global
Metropolis test is proportional to $V^2$, which renders this algorithm
costly for very large lattices. \\

To obtain the low lying spectrum of $Q^2$, we used the Lanczos algorithm.
Attention must be paid to roundoff errors, and we chose the method advocated
by Cullum and Willoughby to filter out spurious ``eigenvalues"
\cite{cw}(see appendix B).

\section{Numerical simulations with the exact algorithm}

Numerical simulations using the exact algorithm described
 in the previous paragraph have been performed on a $4^4$
lattice in the deconfined phase, at $\beta = 6.0$.
For the choice of the ($\epsilon$, $n$)
parameters we have been guided by our experience with the pure
 L\"uscher algorithm. A few exploratory results
 for a $8^4$ lattice, at $\beta=5.3$, and $\kappa=0.156$ and
$0.162$ (therefore, in the confined phase) will be also presented
 in the next section, and the related problems discussed.\\

For the $4^4$ lattice, we concentrated
 on $\kappa = 0.14$, where the
$\epsilon$ range acceptable for the pure L\"uscher algorithm is more
limited.  We explored the capabilities of the modified algorithm
by varying $\epsilon$ and $n$ and
 compared with previous results at the same parameter values.

\begin{table}
\begin{center}
\begin{tabular}{|c|l|l|l|l|l|l|l|}\hline
$\beta$ & $\kappa$ & $\epsilon$ & $n$ & plaquette &
$\delta(n,\epsilon)$ & $accep.$ & $n_{Lan}$\\\hline\hline
 &&&&&&&\\
6.0 &0.14 & 0.02  & 50 & 0.6082(6)  & 1.0$\times 10^{-6}$ & 0.97 & 800 \\
 &     &       & 90 & 0.6081(5)  & 1.0$\times 10^{-11}$ & 0.99 & 800 \\
 &     & 0.04  & 30 & 0.6093(4)  & 7.0$\times 10^{-6}$ & 0.89 & 800 \\
 &     &       & 60 & 0.6084(6)  & 3.6$\times 10^{-11}$ & 0.98 & 800 \\
 &     &       & 90 & 0.6094(5)  & 1.9$\times 10^{-16}$ & 0.98 & 800 \\
 &     & 0.08  & 16 & 0.6072(5)  & 1.0$\times 10^{-4}$ & 0.71 & 300 \\
 &     &       &    & 0.6083(5)  & 1.0$\times 10^{-4}$ & 0.69 & 1000 \\
 &     &       &    & 0.6088(6)  & 1.0$\times 10^{-4}$ & 0.71 & 2000 \\
 & HMC  &       &    & 0.6094(2)  &            &          &    \\\hline
\end{tabular}
\caption{Simulation results for the plaquette on a $4^4$ lattice at
$\beta=6$ for the exact algorithm (L\"uscher+Metropolis)
 and the HMC algorithm. Each
plaquette value is an average of about $10'000$ measurements.}
\end{center}
\end{table}

 The results are
listed in Table 2. In the last column of this table, the number of Lanczos
iterations performed is reported. The relationship between the
number of Lanczos iterations and the actual cut-off $L'$ is linear.
The values for the average acceptance of
 the Metropolis step are also shown.
Finally, since we chose a value of $m$, the number of ($\phi$, $U$) sweeps,
equal to or larger than the autocorrelation time $\tau$ (i.e: $m=10$),
 the autocorrelation of
 the new results in our new units is always $\tau \sim 1$,
and is not shown in the table.\\

In Fig. 4 the plaquette values are plotted vs. $\delta (\epsilon,
n)$, along with the corresponding results from the pure L\"uscher
algorithm.
{}From Table 2 and Fig.  4, it is clear that, when the pure L\"uscher
algorithm gives satisfactory results, acceptance for the Metropolis step
is very high, and
no significant differences are observed
when using the exact algorithm.
Where the approximation of the pure L\"uscher
algorithm breaks down
(in Table 2, for parameters $\epsilon=0.08$, $n=16$),
on the other hand, the average acceptance goes
down to about 70\%, and a significant improvement in the plaquette value
is observed, meaning that the insertion of the $R[U]+1$ term actually
corrects the error the approximation makes for $s< \epsilon$. This is
clearly illustrated in Fig. 5, where the plaquette value is shown to
improve as more and more terms in the product
$\prod_{i=1}^L \lambda_iP(\lambda_i)$ are computed.\\

These results show that the exact algorithm can be effectively used to
perform computations at higher values of $\epsilon$ or lower values of $n$
than acceptable for the pure L\"uscher algorithm,
thus  {\em reducing} the CPU cost of the simulation.\\

As for
the scaling behaviour, that of the original version of the algorithm
is $V(\log V)^2$.
The version of the algorithm with the global Metropolis test introduces
additional work for the evaluation of the low spectrum of the fermion matrix.
Its scaling behaviour is more critical: $aV(\log V)^2~+~bV^2$, where $a\gg b$
are prefactors.

\section{Numerical simulation on the $8^4$ lattice}

We finally present here a few results obtained on an
$8^4$ lattice with the aim of exploring the behaviour of
the pure L\"uscher and exact algorithm in a situation nearer to
 the physical one, i.e. on a larger lattice in the confined phase,
 and with relatively light quarks.
The data taken are not exhaustive, and we only intend
to sketch the possible problems inherent
 to simulations with such parameters, and show the indications
 coming from a few numerical results.\\

The main problem we must be ready to encounter has already been
mentioned: on an $8^4$ lattice, the density of the $Q^2$
eigenvalues is higher than on a $4^4$ one.
This means, on the one hand, that we will
 have to use an $\epsilon$ value much closer to the
 minimum eigenvalue than in the $4^4$ case, and on the other hand that, when
 the global Metropolis test is introduced, a very large number of Lanczos
iterations will be necessary to determine all eigenvalues up to
$\epsilon$.

\begin{table}
\begin{center}
\begin{tabular}{|c|l|l|l|l|l|l|l|l|l|}\hline
k & $\epsilon$ & $n$ & plaquette & $W(1,2)$ & $W(1,3)$ &
 $W(2,2)$  & $\chi(2,2) $ & $accept.$ & $n_{Lan}$\\\hline\hline
 &&&&&&&&&\\
0.156 & 0.007  & 80 & 0.485(1)  & 0.248(1) & 0.128(1) & 0.071(1) &
0.56(3)     &  &  none \\
     &        & 80 & 0.483(1)  & 0.248(3) & 0.127(3) & 0.071(2) &
0.56(3)  &0.88 & 2000 \\
     &        &    & 0.4847(3) & 0.2472(3)& 0.1274(3)&  0.0712(2) &
0.571(8) &   & HMC   \\
0.162& 0.01   & 60 & 0.508(1)  & 0.278(2) & 0.153(2) &  0.096(2) &
 0.46(1)&  & none \\
     &        & 60 & 0.5057(6) & 0.2740(5)& 0.1500(4)& 0.0914(2) &
  0.485(6) & 0.80 & 1000 \\
     &        & 60 & 0.5023(6) & 0.2686(5)& 0.1461(5)& 0.0876(2) &
 0.494(6)  & 0.72 & 2000 \\
     &        &    & 0.5016(4) & 0.2679(5)& 0.1451(4)& 0.0870(4) &
 0.497(6) &   & HMC  \\\hline
\end{tabular}
\caption{Simulation results for the plaquette on a $8^4$ lattice at
$\beta=5.3$ for the L\"uscher+Metropolis
algorithm and the HMC algorithm. Each
plaquette value is an average of about $4'000$ measurements. Here
$W(I,J)$ represents the $I\times J$ Wilson loop, and $\chi(2,2)$
 is the $(2,2)$ Creutz ratio.}
\end{center}
\end{table}

Let us now look, in Table 3 and Fig. 6, at the results of the simulations
 performed at $\beta=5.3$,
$\kappa=0.156$ with $\epsilon=0.007,~ n=80 $ ($\delta=3 \times 10^{-6}$,
lowest eigenvalue
 $\sim 0.001$), and
$\kappa=0.162$ with $\epsilon=0.01,~ n=60 $ ($\delta=1 \times 10^{-6}$,
lowest eigenvalue
 $\sim 0.0005$), both without and with the global Metropolis test.
We chose the $\beta$ and $\kappa$ values such as to reproduce HMC results
by Gupta et al. \cite{gupta}, which are also reported in
Table 3, for comparison.\\

Both the $(\epsilon, n)$ range explored and the statistics are too
 poor to allow us to draw definitive conclusions; nevertheless, we can
derive from these data some indications. For $\kappa=0.156$, where we used
$\epsilon =0.007$, with a lowest eigenvalue $\sim 0.001$, we can see
that the pure L\"uscher algorithm gives quite good results, and the
global Metropolis test
has little influence (corresponding to
 an acceptance rate of 88\%).\\

 When we go to $\kappa=0.162$, with a lowest
eigenvalue $\sim$ 0.0005, and simulate at $\epsilon=0.01$ (20 times
 larger), the L\"uscher algorithm result differs markedly from the
 HMC one; we now introduce the global Metropolis test, and we see that
 it brings the result to agreement with HMC, within
 the statistical error.\\

 One apparently puzzling observation is that
all plaquette and Wilson loop values converge to the HMC ones from
above, whereas, since the quenched results are lower than the full QCD
ones, one might have expected a monotonic convergence from below.
To investigate this phenomenon, we combine the Wilson loop results to form the
$(2,2)$ Creutz ratio $\chi(2,2)$
\be
   \chi(2,2) = -\log{\frac{W(2,2)W(1,1)}{W(1,2)^2}},
\ee
which has a physical significance, since it approximates
$(\sigma a)^2$, the squared string tension in lattice units.
 From Table 3 and Fig. 7 we
see that this quantity converges to the correct limit from below,
whereas the quenched value is above the full QCD one. Thus monotonic
convergence is not guaranteed, even for physical observables. This will
make extrapolation to the exact results a rather delicate matter.\\

{}From the analysis of these
data, we have
confirmation of previous results on the accuracy of the approximation
and also of the expected CPU requirements at larger volume.

\section{Conclusion}
	We have presented first results of full QCD simulations using
L\"uscher's method. Measurements of the plaquette on a $4^4$ lattice
show that the approximation remains quite accurate for that observable,
even when the cutoff parameter $\epsilon$ is increased far above the
lowest eigenvalue $\lambda_{min}(Q^2)$. They also support our analysis of
the complexity of the algorithm: $V (\log V)^2$ with the volume $V$
of the system,
$m_q^{-3}$ or better with the quark mass $m_q$. This compares favorably
with Hybrid Monte Carlo ($V^{5/4}$ and $m_q^{-13/4}$). \\

	Larger Wilson loop measurements on an $8^4$ lattice reveal the
systematic errors of the method. Unlike most approximations to dynamical
quarks, it turns out that the approach of Wilson loops and Creutz
ratios to their correct
values is not monotonic as the approximation improves. \\

	We have tested a variant of L\"uscher's algorithm, which includes
a Metropolis step to correct for the breakdown of the approximation for
small eigenvalues. This extra step brings Wilson loops to their correct
values within errors, for an overhead which is small on our lattice sizes,
but grows like $V^2$.  Further variants are being explored, in particular
an exact non-hermitian one proposed in \cite{borici}.

\section{Acknowledgements}
We acknowledge helpful suggestions from
 A.Bori\c{c}i;
 we thank R.Sommer and B.Jegerlehner for useful
 discussions, and P.Arbenz and M.Gutknecht for insights on the Lanczos
algorithm.
Our thanks also go to all authors of reference \cite{luescher}, for
allowing us access to parts of their software and documentations.
Finally, we thank the SIC of the EPFL for granting us access to the
computer resources of the EPFL.

\newpage
{\Large {\bf Appendix}}\\[1cm]
\appendix
\section{Implementation}
The program is written in FORTRAN 77 for the Cray-T3D using communication
primitives of the Cray. Its structure is organised in four levels and
a user interface facilitates multiuser manipulations. From the bottom up,
the first
level controls communication, navigation (see below) and memory allocation
between the processors (PE's). The second level constructs all objects
needed in the simulations (for example
 staples, plaquettes, $Q$ and $Q^2$ vector
multiplications,...). The third level contains all the algorithms used
for the simulation. The fourth level is the envelope which controls the
whole program. A common user interface to facilitate compilation and
input/output is defined.\\

The lattice $\Lambda$ is first divided into hypercubes of side 2
 (containing $2^4$ points each),
 then adjacent hypercubes are grouped into
 $np$\footnote{Here $np$ represents the number of used PE's. On
the Cray-T3D only $np$ equal to a power of 2 is allowed.}
  equal sets $\Lambda_i\subset\Lambda~~~(i=0,...,np-1)$,
 each of which will be assigned to one PE.
 For the navigation in the lattice
a mapping $Navig: \Lambda\rightarrow \Lambda_i$
between
the physical coordinates on the lattice and their allocation as
local sublattice coordinates is stored as a
table on each PE. For a given lattice coordinate $x\in\Lambda$
the mapping $Navig$ returns the corresponding local coordinate
$x_i\in\Lambda_i$ and the identification number $i$ of the PE which
owns $x_i$. The memory allocation of gauge and bosons fields is
organised according to this mapping. The communication of data is also
controlled by this mapping.\\

All objects are constructed in parallel by the PE's which locally
operate on their sublattices. The
construction of the objects needs some communication of data between PE's.
The communication is performed asynchronously using the
$shmem\_get/shmem\_put$ routines of the Cray-T3D.
Only at the algorithmic level the needed synchronisations are defined.\\

All algorithms used (heat-bath, over-relaxation, global
Metropolis, Lanczos) are constructed using as input the objects.
The heat-bath and over-relaxation algorithms for the boson fields
are specifically written for the L\"uscher local bosonic action. All
other algorithms are general and can be also used for quenched
simulations by simply switching some libraries at level 2.\\

\section{Lanczos process}

The Lanczos algorithm is a powerful technique
used for the determination of the spectrum of
sparse, symmetric matrices\cite{lanz}. During the iterations of the
Lanczos algorithm a $m\times m$ tridiagonal symmetric matrix $T^{(m)}$
is obtained by transforming a $n\times n$ matrix $A$ (with $m\ll n$)
with a matrix $Q$ whose columns are called Lanczos vectors.
The extremal eigenvalues of $T^{(m)}$ are estimates of the extremal
eigenvalues of $A$.

In exact arithmetic, the Lanczos vectors are orthogonal.
However, in finite precision arithmetic, they lose mutual
orthogonality as the number of iteration steps $m$ increases.
As a consequence eigenvalues re-appear, i.e. the algorithm
finds ``spurious'' or ``ghost'' values which are not actual eigenvalues of $A$,
but which are nearly degenerate with them.\\

 The problem can be solved by explicitly
orthogonalizing the Lanczos vectors.
However, all the Lanczos vectors must then be stored and the memory
cost of the algorithm is enormous.\\

Instead we use another clever way of identifying spurious eigenvalues without
orthogonalization, proposed by Cullum and Willoughby \cite{cw}. In
their algorithm one the eigenvalues of $T^{(m)}$ are compared with the
eigenvalues of a matrix $T_2$ which is obtained from $T^{(m)}$
by deleting its first row and column. If a simple eigenvalue of $T^{(m)}$
is also a simple eigenvalue of $T_2$, then this eigenvalue is spurious.\\

\newpage
{\bf {\Large Figure caption}}
\begin{enumerate}
\item  Lower spectrum of the $Q^2$ matrix at $\beta=6$ and $\kappa=0.12$
on a $4^4$ lattice.

\item Plaquette versus $\delta$ at $\beta=6$, (a) $\kappa=0.12$ and (b)
$\kappa=0.14$ on a $4^4$ lattice with
the pure L\"uscher algorithm.
The lines indicate the HMC values, with their errors.

\item Autocorrelation time of the plaquette versus $n/\sqrt{\epsilon}$
for a $4^4$ lattice, at $\beta=6.$, (a) $\kappa=0.12$ and (b)
$\kappa=0.14$. The lines represent linear fits to the data.

\item Plaquette versus $\delta$ at $\beta=6$,
$\kappa=0.14$ on a $4^4$ lattice with the use of the global Metropolis test.
The lines indicate the HMC values, with their errors.

\item Plaquette versus number of Lanczos iterations used in the global
 Metropolis test, at $\beta=6$, $\kappa=0.14$ on a $4^4$ lattice,
 with parameters $\epsilon=0.08$ and $n=$16. The lines indicate the HMC value,
 with its error.

\item Plaquette (a) and Wilson loop $WL(1,2)$ (b) data
 from the simulation on the $8^4$
lattice at $\beta=5.3$ with $\kappa=0.156$ ($\Diamond$)
 and $\kappa=0.162$ ($\Box$).
 The lines indicate the
 HMC values, with their errors.

\item Creutz ratio $\chi(2,2)$ versus number of Lanczos
iterations used in the global Metropolis test, on the $8^4$ lattice,
$\beta=5.3$, $\kappa=0.162$, $\epsilon=0.01$ and $n=60$.
The lines indicate the HMC value, with its error.
\end{enumerate}

\newpage

\setlength{\unitlength}{0.240900pt}
\ifx\plotpoint\undefined\newsavebox{\plotpoint}\fi
\sbox{\plotpoint}{\rule[-0.200pt]{0.400pt}{0.400pt}}%


\vspace{3cm}

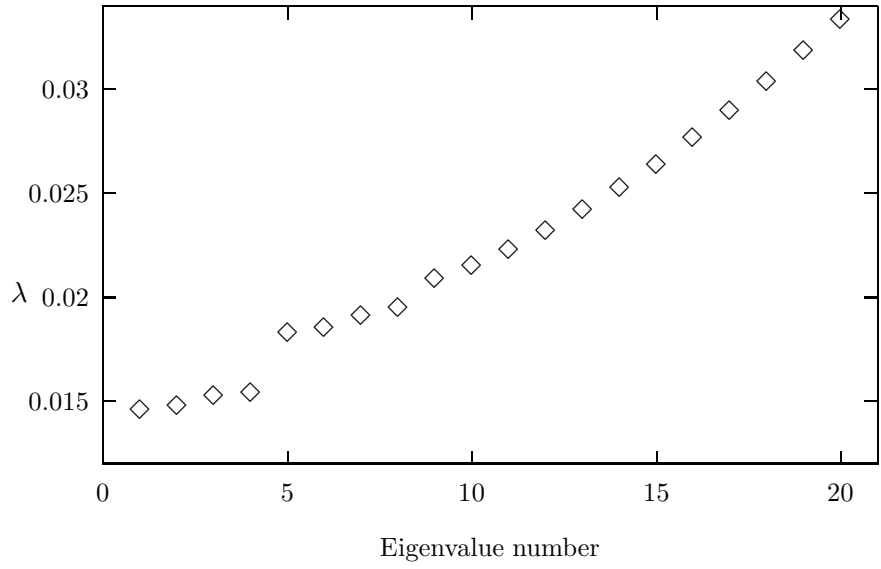
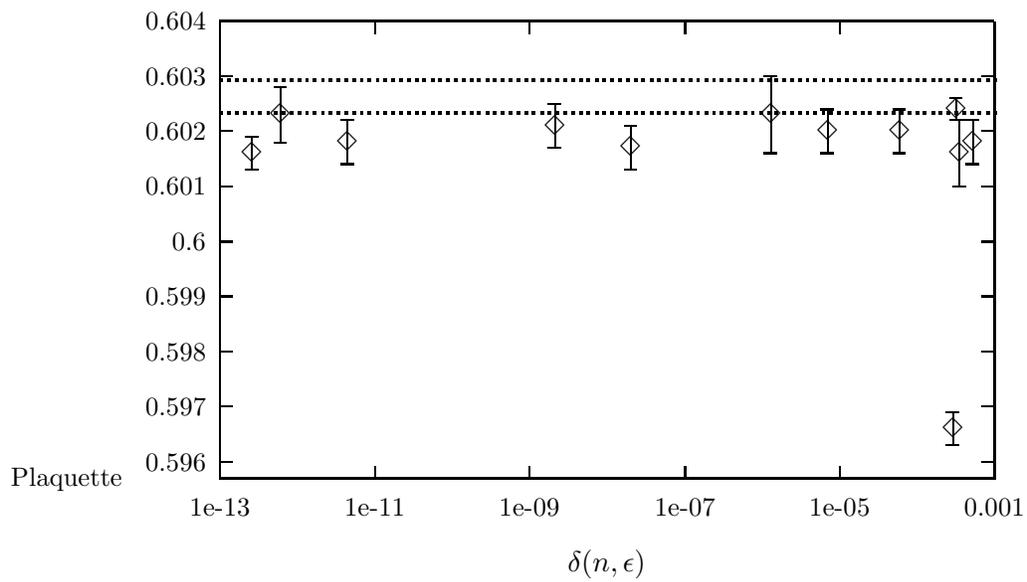
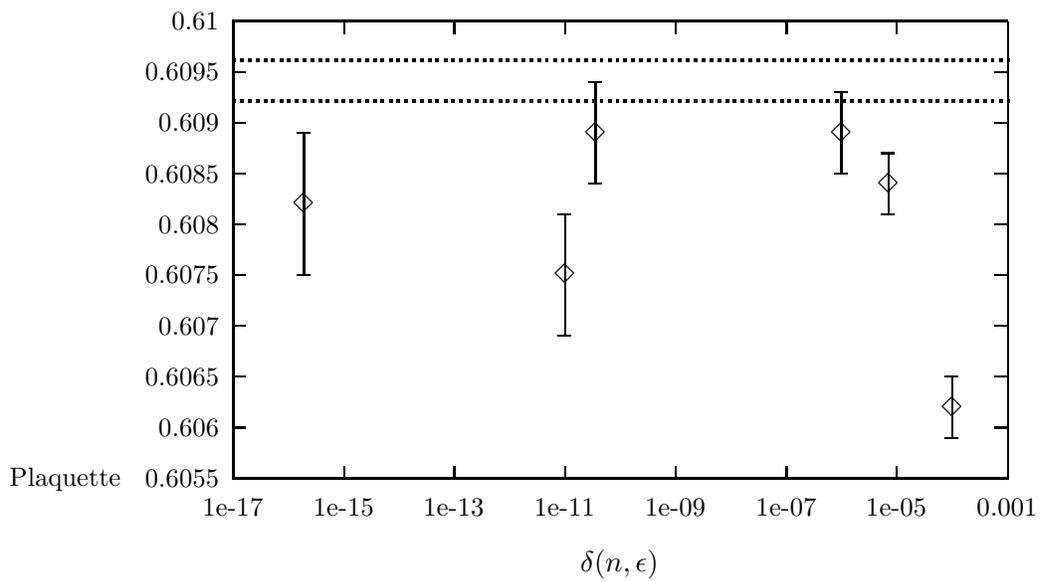
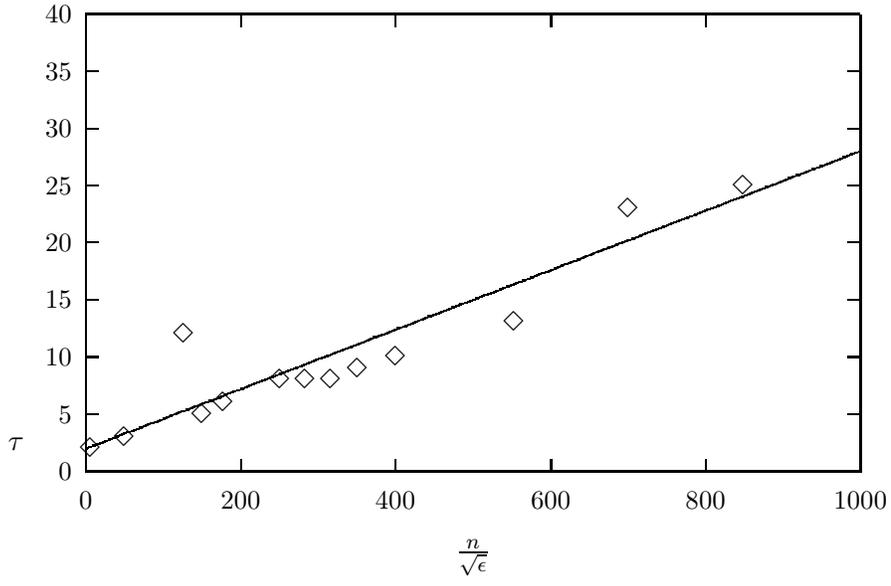
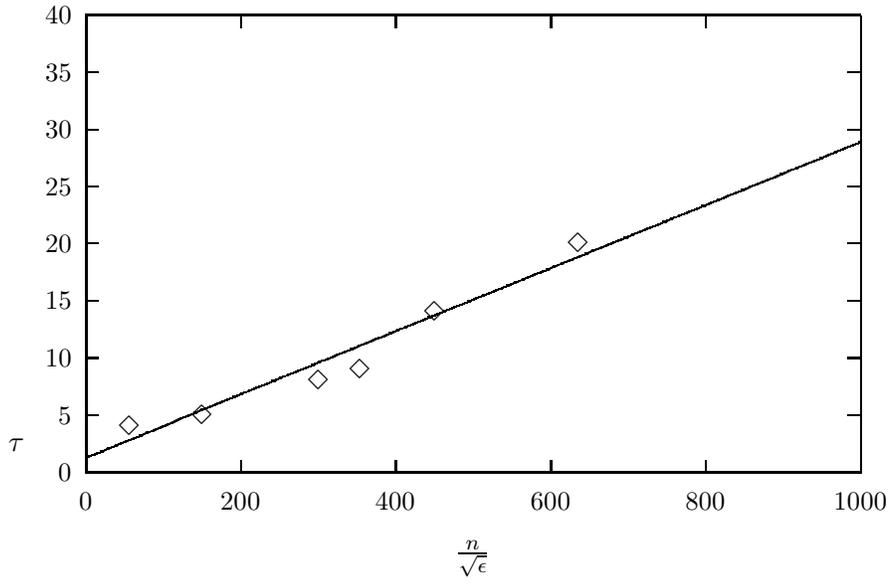
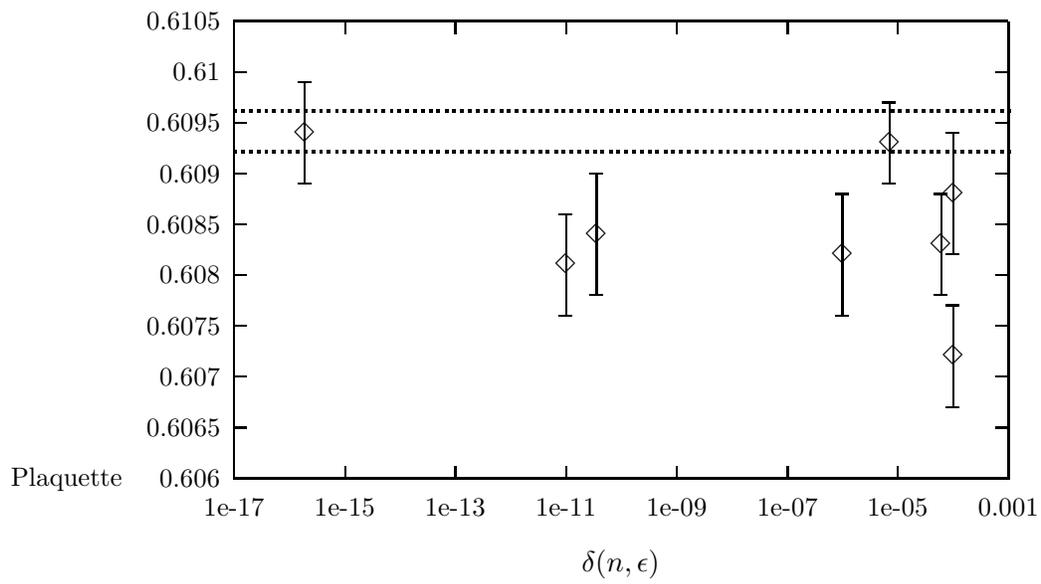
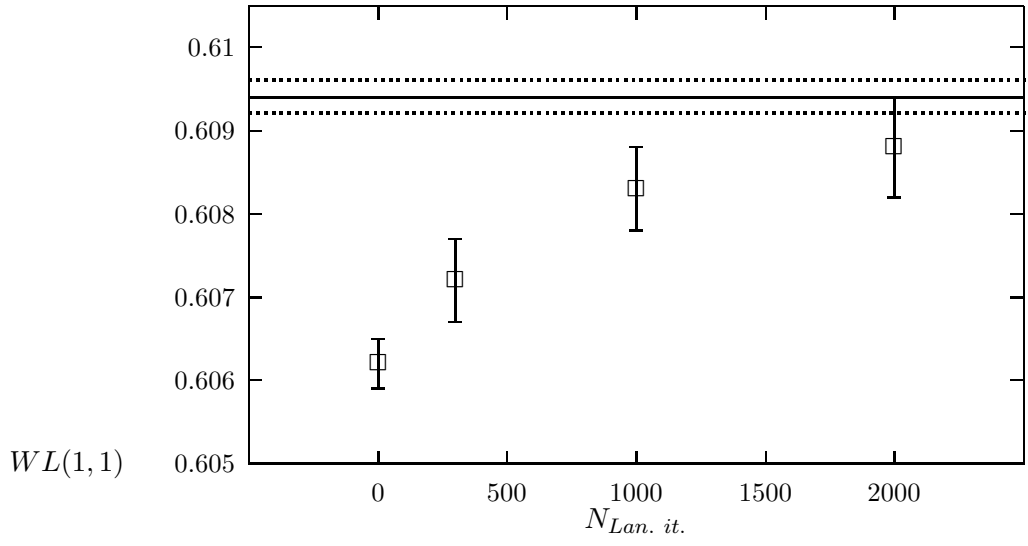
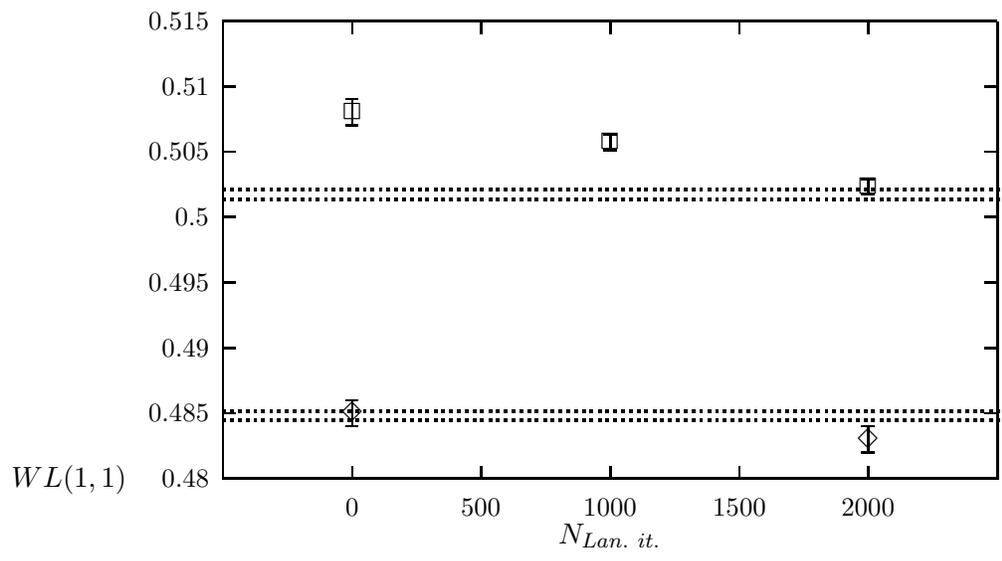
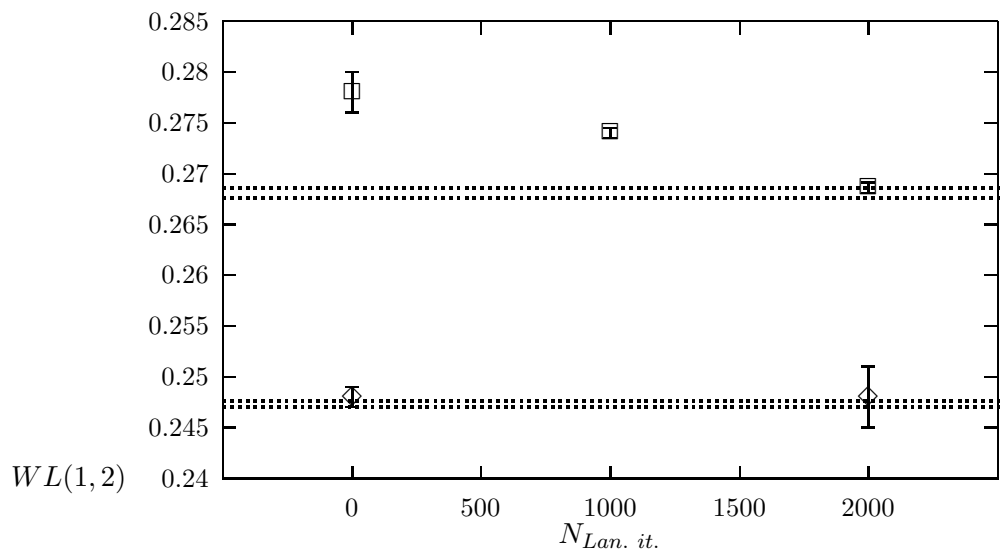
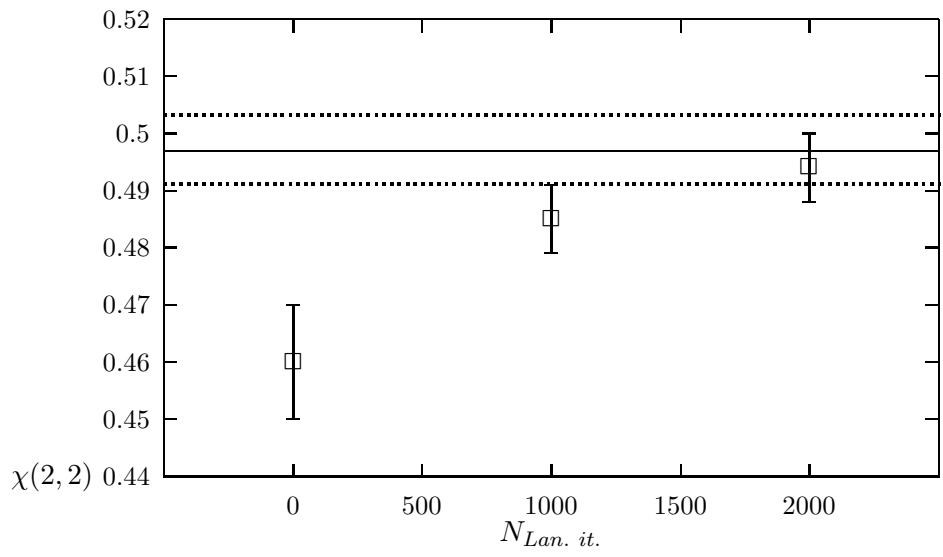
\end{document}